# Structural instability of EuTiO$_3$ from X-ray powder diffraction

J. Köhler, R. Dinnebier, A. Bussmann-Holder

Max-Planck-Institute for Solid State Research, Heisenbergstr. 1, D-70569 Stuttgart, Germany

We have recently predicted and subsequently verified experimentally by specific heat measurements that EuTiO$_3$ undergoes a structural phase transition at elevated temperature $T_S$ = 282 K. The origin of the phase transition has been attributed to the softening of a transverse acoustic mode stemming from an oxygen octahedral rotation analogous to SrTiO$_3$. Here we demonstrate that the theoretical interpretation is correct by using high resolution laboratory X-ray powder diffraction which evidences a cubic to tetragonal phase transition in EuTiO$_3$. The room temperature structure could be refined in $Pm\bar{3}m$ with $a$ = 3.9082(2) Å and at 100 K the refinement in the tetragonal space group $I4/mcm$ resulted in $a$ = 5.5192(2) and $c$ = 7.8164(8) Å.

Perovskite titanates are well investigated due to their complexity in their phase diagram and their dielectric and ferroelectric properties. In spite of the fact that ferroelectricity is absent in SrTiO$_3$ (STO), it is a quantum paraelectric in which the transition towards a ferroelectric state is suppressed by quantum fluctuations [1], it is one of the best investigated titanates. STO serves as a substrate in thin film techniques [2], invokes novel interface properties in layered materials [3, 4], exhibits unusual elastic properties [5], and can become ferroelectric upon oxygen isotope replacement [6, 7]. Furthermore, it has huge dielectric constants at cryogenic temperatures [8], shows precursor dynamics to its displacive structural transition at $T_S$=105 K [9], and exhibits unusual features around 40 K as evidenced by ultrasonic attenuation [10]. As such it is worthwhile to study a system with similar properties and eventually increased functionalities as, e.g., offered by EuTiO$_3$ (ETO). ETO is very similar to STO since Eu and Sr have the valency +2 and almost the same ionic radii, which implies that also their lattice parameters are nearly the same. In strong similarity to STO, ETO also shows a pronounced long wave length optic mode softening, indicating its tendency towards ferroelectricity [11 – 13], but being suppressed by quantum fluctuations as in STO.

ETO was synthesized already in 1966 [14] in the search for other ferroelectric perovskites, but attracted only more attention rather recently when it was shown that the low temperature phase transition to an antiferromagnetic state at $T_N$ = 5.5 K has a substantial influence on its dielectric constant [11 – 13]. At $T_N$ the dielectric constant experiences a rapid drop, which can be reversed by the application of a magnetic field. This effect suggests strong spin-lattice coupling and an inherent tendency towards multiferroic properties. This enlarged multifunctionality of ETO as compared to STO is expected to become important in all aspects mentioned above for STO and tunes various layer, substrate and interlayer properties. Especially, it should be possible to trigger the magnetic properties of ETO through interfacial coupling.

Recently we have predicted a further analogy between STO and ETO [15], namely the existence of an oxygen ion octahedral rotation instability at elevated temperatures which was confirmed experimentally indirectly by specific heat measurements [15]. The comparison of the specific heat anomalies seen at the structural transition temperatures of STO and ETO suggests that they are of the same origin. However, details about the low temperature structure and the space group of ETO were not given at that time. Only from theoretical considerations the low temperature phase has been assigned as tetragonal [15, 16] with the same space group as STO. This implies that a zone boundary transverse acoustic mode frequency softens with decreasing temperature to become unstable at $T_S$=282K.

The structural refinement of the low temperature phase of ETO is expected to be similarly difficult as the one in STO where the tetragonal lattice parameter ratio is as small as $c/a$=1.00056. For this system initially only ESR, EPR, and INS [1, 17 – 21] were able to see the phase transition and to allow conclusions about the tilting instability of the oxygen octahedra. Later also specific heat measurements detected a tiny anomaly at $T_S$ in STO [22 – 25]. The space group of the low temperature phase of STO was indirectly assigned by ESR as $I4/mcm$ [26] and later confirmed by X-ray crystal structure analysis [27]. While in a variety of earlier X-ray work on STO [28] only the lattice parameters of the low temperature phase were determined, the structural refinement of Ref. 27 concentrated on the detection of the very weak superlattice reflections caused by the multiplication of the unit cell to √2a×√2a×2a, with *a* being the cubic lattice parameter. Similar difficulties in the structure refinement are expected for ETO where the newly discovered phase transition has not been detected before our recent report [15].

In order clarify the structure of ETO below $T_S$=282K, ETO samples were prepared by carefully mixing dried $Eu_2O_3$ ((Alfa, 99.99%) with $Ti_2O_3$ powder (Alfa, 99.99%), in a 1:1 ratio in an agate mortar under Ar. Then the powder was pressed to a pellet, and heated in a corundum tube under Ar for 4d at 1400°C. The ETO sample was dark grey and cubic at room temperature according to X-ray powder diffraction data.

Subsequently, high resolution X–ray powder diffraction data of $EuTiO_3$ were collected at -180°C and 25°C on a laboratory powder diffractometer Bruker-D8–Advance (Cu–$K_{\alpha 1}$ radiation from a primary Ge(111); Våntec position–sensitive detector) in Bragg-Brentano geometry using a TC - WIDE RANGE chamber (MRI GmbH, Karlsruhe). Data at both temperatures were taken in steps of 0.016° 2θ from 5.0 – 115.0° 2θ for 24 hrs.

For the structure determination and refinement of the low temperature phase, the program TOPAS 4.1 [29] was used. Indexing of the diffraction data of $EuTiO_3$ at -180°C was performed by iterative use of singular value decomposition as implemented in the program TOPAS [30], leading first to a primitive tetragonal unit cell (Table 1). The unit cell of the low temperature structure of ETO could be refined in *P4mm* with *a* = 3.9027(2) Å and *c* = 3.9082(2) Å under the inclusion of all observed reflections. However, Rietveld refinements [31] showed that in this structural model no minimum for the x-parameter of O(2) could be found, with an estimated standard deviation approximately one order of magnitude larger than the expected deviation from the special position. This means that from the intensity profile (Figure 1) a different setting of the unit cell and space group is required. In analogy to STO [27] this setting is based along the diagonals within the *ab*-plane and a doubling along the tetragonal *c*-axis corresponding to the space group $I4/mcm$. From the volume increment Z was determined to be 4. The peak profile and precise lattice parameters were determined by Le Bail fits [31] using the fundamental parameter (FP) approach of TOPAS [31]. For the modeling of the background, Chebychev polynomials were employed. The refinement converged quickly.

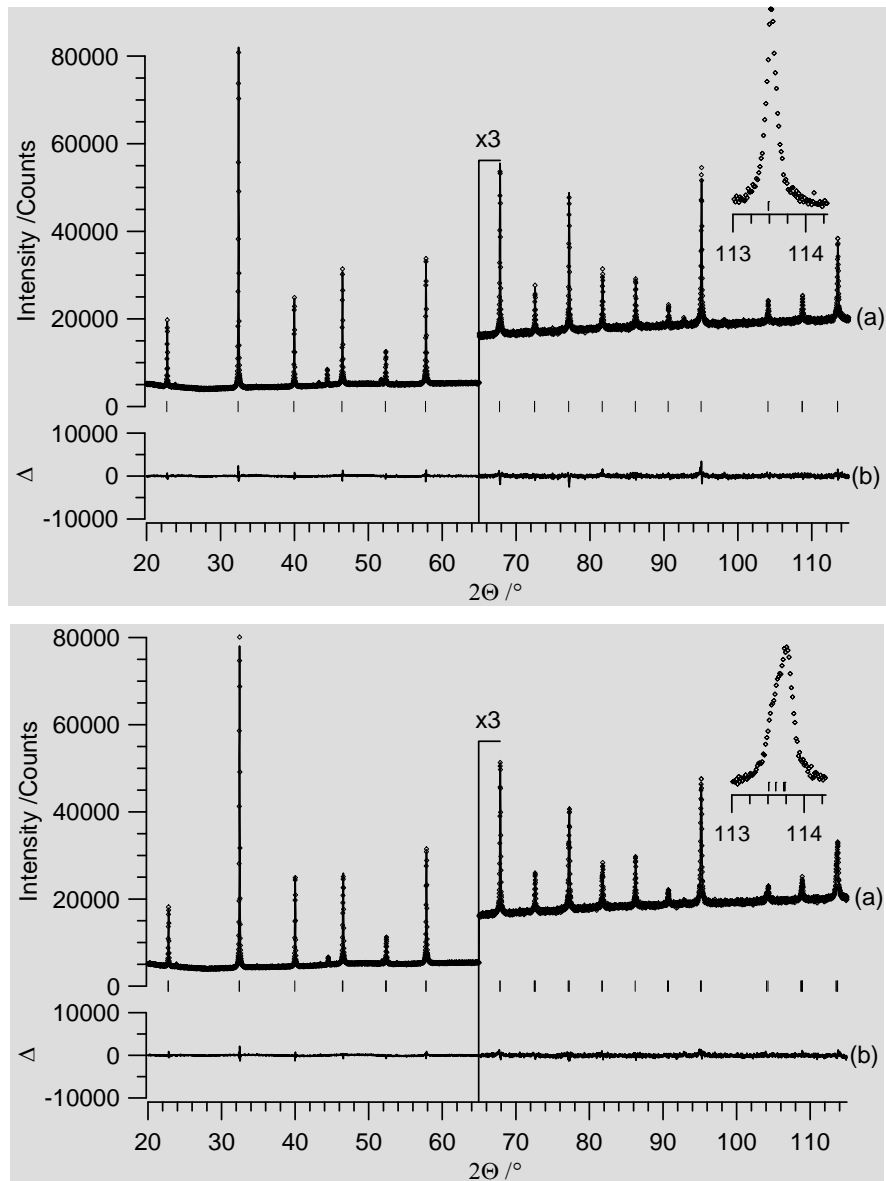

**Figure 1** Scattered X-ray intensity for EuTiO$_3$ at 30°C (top) and -180°C (bottom) as a function of diffraction angle 2θ. The observed pattern (diamonds), the best Rietveld-fit profile (line a), the difference curve between observed and calculated profile (line b), and the reflection markers (vertical bars) are shown. The wavelength was λ = 1.54059 Å. The higher angle part of the plot starting at 65°2θ is enlarged for clarity. The inset shows the cubic (411)/(033) reflection (top) and its tetragonal splitting into 4 reflections (bottom).

Atomic starting parameters for Rietveld refinement of the room temperature crystal structure of EuTiO$_3$ were taken from Brous *et al.* [32], while for the low temperature crystal structure of EuTiO$_3$ those of tetragonal BaTiO$_3$ and PbTiO$_3$ [33] were used. All profile and lattice parameters were released and all atomic positions were subject to free unconstrained refinement. The refinement converged quickly. Final agreement factors (R-values) are listed in Table 1 for the high and low temperature structures together with lattice parameters. We have added, for comparison, the same parameters for the low temperature structure in cubic notation.

**Table Captions:** Crystallographic data for EuTiO$_3$.
**Table 1**.

| Formula | EuTiO$_3$ | EuTiO$_3$ | EuTiO$_3$ |
|---|---|---|---|
| Temperature (C) | 30 | -180 | -180 |
| Space group | $Pm\bar{3}m$ | $I4/mcm$ | in $P4mm$ notation |
| Z | 1 | 4 | 1 |
| a (Å) | 3.9082(2) | 5.5192(2) | 3.9027(2) |
| c (Å) | 3.9082(2) | 7.8164(8) | 3.9082(2) |
| V (Å$^3$) | 59.693(8) | 238.1042(8) | 59.527(7) |
| δ-calc (g/cm$^3$) | 6.895(1) | 6.895(1) | 6.895(1) |
| R-p (%)[a] | 1.48 | 1.45 | 1.45 |
| R-wp (%)[a] | 1.94 | 1.88 | 1.88 |
| R-F$^2$ (%)[a] | 0.92 | 0.85 | 0.85 |
| GoF[a] | 1.48 | 1.44 | 1.44 |
| Starting angle (°2θ) | 5.0 | 5.0 | 5.0 |
| Final angle (°2θ) | 115.0 | 115.0 | 115.0 |
| Step width (°2θ) | 0.016 | 0.016 | 0.016 |
| Time/scan (hrs) | 24 | 24 | 24 |

[a] R-exp, R-p, R-wp, R-F$^2$, and GoF as defined in TOPAS (Bruker AXS)

Further details of the crystal structure investigations may be obtained from Fachinformationszentrum Karlsruhe, 76344 Eggenstein-Leopoldshafen, Germany (fax:

(+49)7247-808-666; e-mail: crysdata(at)fiz-karlsruhe.de, http://www.fizkarlsruhe.de/request_for_deposited_data.html) on quoting the appropriate CSD number xxxxxx.

A schematic representation of the low temperature structure is presented in Figure 2 where for clarity the oxygen octahedral rotation angle has been enlarged.

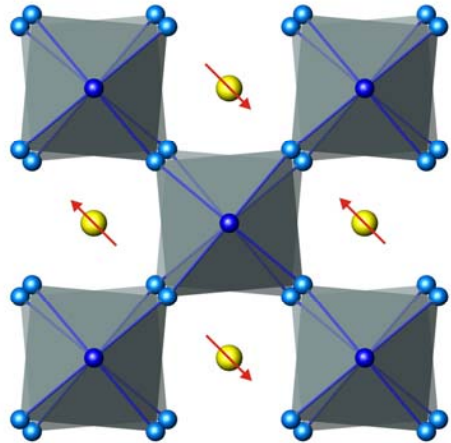

**Figure 2** The schematic crystal structure of EuTiO$_3$ at -180°C in a projection along the *b*-axis, exhibiting TiO$_6$ polyhedra (blue shaded). The oxygen ions are displayed in blue, the Eu ions in yellow. The arrows on the Eu ions indicate the low temperature spin structure.

In order to compare the structural refinement of ETO with the one of STO, the cubic *c*/*a* ratio is readily obtained from the above data. While this ratio is 1.00062 in STO at 4.2K [26], i.e. 100 K below the transition temperature, it is 1.0014 in ETO at 93 K, almost 200 K below the phase transition. This ratio can be related to the angle of rotation $\varphi$ of the oxygen octahedron via the relation $c/a = 1/\cos\varphi$ [26]. While in STO $\varphi = 2.1°$ at 4.2 K, in ETO it is $\varphi = 3.03°$ at 93 K. Since $\varphi$ is the order parameter of the phase transition [33], its squared value varies linearly with t= T/T$_S$ as long as T is not too close to T$_S$. By using this relation and comparing the value of $\varphi$ of ETO at 93K with data of STO, the extrapolated zero temperature value of the rotation angle is estimated to be 3.37 which is rather large as compared to STO, but a consequence of the fact that T$_S$ of ETO is much larger than T$_S$ of STO.

This latter conclusion needs to be verified by temperature dependent structural data. The proposed oxygen ion rotational instability can further be confirmed by inelastic neutron scattering experiments. An additional support is also expected to come from ESR, EPR and Mössbauer measurements where a line splitting should appear at 282(1)K.

In summary, a novel phase transition has been predicted to exist in ETO [15, 16], analogous to that in STO, wherein the oxygen octahedra tilt at the theoretical transition temperature $T_S \approx 298K$. Experimentally this transition has been detected at $T_S=282$ K [15] and tentatively ascribed to the same transition as observed in STO, namely an oxygen ion rotational instability caused by the softening of the zone boundary acoustic mode. Here we have shown that indeed the low temperature structure is tetragonal confirming theory. The structural refinement assigned the low temperature phase to the space group $I4/mcm$ with a zero temperature rotation angle of 3.37 – almost 50% larger than in STO. Similarly, the *c/a* cubic lattice constant ratio is enlarged where both of these enhancements are attributed to the fact that $T_S$ of ETO is more than twice as large as $T_S$ of STO.

________

**References**

1. K. A. Müller and H. Burkhard, Phys. Rev. B **19**, 3593 (1979).
2. see e.g. Masashi Kawasaki, Kazuhiro Takahashi, Tatsuro Maeda, Ryuta Tsuchiya, Makoto Shinohara, Osamu Ishiyama, Takuzo Yonezawa, Mamoru Yoshimoto and Hideomi Koinuma, Science **266**, 1540 (1994).
3. A. D. Caviglia, S. Gariglio, N. Jaccard, T. Schneider, M. Gabay, S. Thiel, H. Hammerl, J. Mannhart, J. M. Triscone, Nature **456**, 624 (2008).
4. Guneeta Singh-Bhalla, Christopher Bell, Jayakanth Ravichandran, Wolter Siemons, Yasuyuki Hikita, Sayeef Salahuddin, Arthur F. Hebard, Harold Y. Hwang & Ramamoorthy Ramesh, Nature Physics **7**, 80 (2011).
5. Edward Poindexter and A. A. Giardini, Phys. Rev. **110**, 1068 (1958).
6. M. Itoh, R. Wang, Y. Inaguma, T. Yamaguchi Y-J. Shan, and T. Nakamura, Phys. Rev. Lett. **82**, 3540 (1999).
7. A. Bussmann-Holder, H. Büttner, and A. R. Bishop, J. Phys.: Cond. Mat. **12**, L115 (2000).

8.  A. S. Mischenko, Q. Zhang, J. F. Scott, R. W. Whatmore, and N. D. Mathur, Appl. Phys. Lett. **89**, 242912 (2006).
9.  Th. Von Waldkirch, K. A. Müller, and W. Berlinger, Phys. Rev. B **7**, 1052 (1973).
10. Chen An, J. F. Scott, Zhi Yu, H. Ledbetter, and J. L. Baptista, Phys. Rev. B **59**, 6661 (1999).
11. S. Kamba, D. Nuzhnyy, P. Vaněk, M. Savinov, K. Knížek, Z. Shen, E. Šantavá, K. Maca, M. Sadowski, and J. Petzelt, Europhys. Lett. **80**, 27002 (2007).
12. V. Goian, S. Kamba, J. Hlinka, P. Vaněk, A. A. Belik, T. Kolodiazhnyi, and J. Petzelt, Eur. Phys, J. B **71**, 429 (2009).
13. T. Katsufuji and H. Takagi, Phys. Rev. B **64**, 054415 (2001).
14. T. R. McGuire, M. W. Shafer, R. J. Joenk, H. A. Halperin, and S. J. Pickart, J. Appl. Phys. **37**, 981 (1966).
15. A. Bussmann-Holder, J. Köhler, R. K. Kremer, and J. M. Law, Phys. Rev. B **83**, 212102 (2011).
16. Jerry L. Bettis, Myung-Hwan Whangbo, Jürgen Köhler, Annette Bussmann-Holder, and A. R. Bishop, Phys. Rev. B **84**, 184114 (2011).
17. K. A. Müller, Phys. Rev. Lett. **2**, 341 (1959).
18. E. J. Kirkpatrick, K. A. Müller, and R. S. Rubins, Phys. Rev. **135**, A86 (1964).
19. G. Shirane and Y. Yamada, Phys. Rev. **177**, 858 (1969).
20. R. A. Cowley, W. J. L. Buyers, and G. Dolling, Solid St. Comm. **7**, 181 (1969).
21. J. M. Worlock, J. F. Scott, and P. A. Fleury, in "Light scattering analysis of Solids", ed. G. B. Wright (Springer, New York, 1969) p. 689.
22. V. Franke and E. Hegenbarth, Phys. Sta. Sol. **25**, K17 (1974).
23. I. Hatta, Y. Shiroishi, K. A. Müller, and W. Berlinger, Phys. Rev. B **16**, 1138 (1977).
24. M. C. Gallardo, R. Burriel, F. J. Romero, F. J. Gutiérrez, and E. K. H. Salje, J. Phys.: Cond. Mat. **14**, 1881 (2002).
25. E. K. H. Salje, M. C. Gallardo, J. Jiménez, F. J. Romero, and J. Del Cerro, J. Phys.: Condens. Matter 10, 5535 (1998).
26. H. Unoki and T. Sakudo, J. Phys. Soc. Jpn. **23**, 546 (1967).
27. H. Fujishita, Y. Shiozaki, and E. Sawaguchi, J. Phys. Soc. Jpn. **46**, 581 (1979).

28. F. W. Lytle, J. Appl. Phys. **35**, 2212 (1964).

29. Bruker AXS, Topas, version 4.1. 2007.

30. A. A. Coelho, J. Appl. Cryst. **36**, 86 (2003).

31. H. M. Rietveld, J. Appl. Cryst. **2**, 65 (1969)..

32. J. Brous, I. Fankuchen, and E. Banks, Acta Cryst. **6**, 67 (1953).

33. K. A. Müller and W. Berlinger, Phys. Rev. Lett. **26**, 13 (1971).